# Vagal contributions to fetal heart rate variability: an omics approach

Christophe L. Herry[2*], Patrick Burns[1*], André Desrochers[1], Gilles Fecteau[1], Lucien Daniel Durosier[3], Mingju Cao[3], Andrew JE Seely[2] and Martin G. Frasch[3,4,5]

[1] Clinical Sciences, CHUV, l'Université de Montréal, St-Hyacinthe, QC, Canada
[2] Ottawa Hospital Research Institute, University of Ottawa, ON, Canada
[3] Departments of Obstetrics and Gynaecology and Neurosciences, CHU Ste-Justine Research Centre, l'Université de Montréal, QC, Canada
[4] Centre de recherche en reproduction animale (CRRA), l'Université de Montréal, St-Hyacinthe, QC, Canada
[5] Department of Obstetrics and Gynecology, University of Washington, Seattle, WA, USA

*PB and CH contributed equally to this manuscript.

**Corresponding author:**
Martin G. Frasch
Department of Obstetrics and Gynecology
University of Washington
1959 NE Pacific St
Box 356460
Seattle, WA 98195
Phone: +1-206-543-5892
Fax: +1-206-543-3915
Email: mfrasch@uw.edu

**Keywords**: In utero, fetal surgery, vagus, autonomic nervous system, HRV

**Abstract**

Fetal heart rate variability (fHRV) is an important indicator of health and disease, yet its physiological origins, neural contributions in particular, are not well understood. We aimed to develop novel experimental and data analytical approaches to identify fHRV measures reflecting the vagus nerve contributions to fHRV. In near-term ovine fetuses, a comprehensive set of 46 fHRV measures was computed from fetal pre-cordial electrocardiogram recorded during surgery and 72 hours later without (n=24) and with intra-surgical bilateral cervical vagotomy (n=15). The fetal heart rate did not change due to vagotomy. We identify fHRV measures specific to the vagal modulation of fHRV: Multiscale time irreversibility asymmetry index (AsymI), Detrended fluctuation analysis (DFA) $\alpha_1$, Kullback-Leibler permutation entropy (KLPE) and Scale dependent Lyapunov exponent slope (SDLE α). We provide a systematic delineation of vagal contributions to fHRV across signal-analytical domains which should be relevant for the emerging field of bioelectronic medicine and the deciphering of the "vagus code". Our findings also have clinical significance for *in utero* monitoring of fetal health during surgery.

**Key points**

- Fetal surgery causes a complex pattern of changes in heart rate variability measures with overall reduction of complexity or variability
- At 72 hours after surgery, many of the HRV measures recover and this recovery is delayed by an intrasurgical cervical bilateral vagotomy
- We identify HRV pattern representing complete vagal withdrawal that can be understood as part of "HRV code", rather than any single HRV measure
- We identify HRV biomarkers of recovery from fetal surgery and discuss the effect of anticholinergic medication on this recovery

**Introduction**

Variability is defined as patterns of fluctuation of a data series occurring over intervals in time. Various aspects of variability can be measured mathematically in several complementary domains, and hundreds of techniques of variability analysis have been proposed in the medical and broader literature.(Electrophysiology, Task Force of the European Society of Cardiology & the North American Society of Pacing, and, 1996; Bravi *et al.*, 2011; Sassi *et al.*, 2015) While some of these techniques capture properties of time series' variability that are correlated, most offer a unique aspect of variability by characterizing different mathematical properties of the signals. They include measures characterizing the statistical properties (e.g., standard deviation, root mean square of successive differences of R-R intervals of electrocardiogram (ECG), RMSSD), the informational complexity (e.g., entropy measures), the pattern of variations across time scales (i.e., time-invariant features such as fractal measures, power law exponents) or the energy contained in the signal (e.g., spectral measures). Researchers generally agree that a plurality of techniques offers the most complete evaluation.(Goldberger, 1996; Goldberger *et al.*, 2002)

An important underlying assumption about studying biological variability is that it has predictive properties about physiological systems from which it emerges. Hence, variability properties are a form of biological code, taking place on different scales in space and time. While the time component can be taken literally, the space component can represent different organs and regions of the body contributing to the properties of a variability pattern in question. The space component can also represent state space in terms of physical signal properties. The complexity of variability encompasses the regularity, degree of information and scale-invariant, self-similar fractal properties of physiological signals and offers a window into the temporal and spatial changes in structure and connectivity of complex dissipative systems. The degree of variability represents the amount of variation and relates to the physiological adaptability of the body, for instance the ability to augment or decrease work as required by external demands.(Carrillo *et al.*, 2016; Wang *et al.*, 2018)

Here we are concerned with a particular kind of biological variability characterized by complex, dynamic spatiotemporal patterns, the heart rate variability (HRV). HRV is an important indicator of health and disease and decreased variability is associated with age and illness and correlates with illness severity, indicating reduced adaptability and/or increased stress.(Glass, 2001; Seely & Macklem, 2004; Macklem, 2008; Macklem & Seely, 2010; Leicht *et al.*, 2018) The physiologic interpretation of variability in time series and HRV in particular continues to be actively explored, impeded by the fact that HRV represents a complex, nonlinear, integration of activities of the sympathetic and parasympathetic nervous systems, influenced by body and breathing movements, baroreflex and circadian processes as well as various, quantitatively poorly understood homeokinetic brain-body communication processes, such as sensing and regulation of cytokine levels or metabolism.(Thayer & Sternberg, 2006; Tomokazu *et al.*, 2009; Jensen *et al.*, 2009; Durosier *et al.*, 2015; Liu *et al.*, 2016; Zanos *et al.*, 2018). In HRV, we find features of spatial variability represented by contributions from various organs and systems such as heart, lung, brain, vasculature and endocrine system. We also find clear characteristics of state space variability in the above sense as well as rich temporal and time-invariant, fractal dynamics. Much like the established notion of neural code and the ongoing attempts to crack it (Borst & Theunissen, 1999; Birmingham *et al.*, 2014; Bouton, 2017), the notion of HRV code seems hence inevitable and logical to introduce and pursue.

While early HRV literature focused on the link between time-domain (e.g., RMSSD) and frequency domain (LF, HF) HRV measures and autonomic modulation (Electrophysiology, Task Force of the European Society of Cardiology & the North American Society of Pacing, and, 1996; Montano *et al.*, 1998), it is clear that estimating autonomic activity with these simple linear metrics is limited and complicated by the previously mentioned nonlinear interconnection of complex sub-systems and processes.(Hopf *et al.*, 1995; Skinner *et al.*, 2000; Frasch *et al.*, 2009; Billman, 2011) Another approach promoted by Seely *et al.* is to try and quantify the degree and complexity of variability as separate, yet complementary coupled concepts.(Seely & Macklem, 2012)

It is commonly held that HRV provides window into vagal control: the effects of vagal modulation on some linear and nonlinear HRV properties have been reported.(Tulppo *et al.*, 2001, 2005; Cerutti *et al.*, 2009; Frasch *et al.*, 2009) The assessment of sympathetic contributions has remained elusive, in part at least due to the complex overlap of vagal and sympathetic oscillations in the low frequency/long term time scales, albeit some studies indicated that complexity properties of HRV can reveal influences specific to sympathetic modulations.(Frasch *et al.*, 2009)

While in adult physiology and medicine, ECG- or blood pressure-derived HRV represents one of many non-invasive monitoring modalities of health and disease, in fetal monitoring, HRV, derived from maternal abdominal ECG or from cardiac ultrasound, provides the only non-invasive window into fetal wellbeing or early detection of disease.(Frasch, 2018)

While attempts to assign some fHRV properties to vagal contributions have been made over the years (Karin *et al.*, 1993; Groome *et al.*, 1994; Ramaekers *et al.*, 2002; Thayer & Sternberg, 2006; Frasch *et al.*, 2007, 2009; Fairchild *et al.*, 2011), it is not known how the complex communication via the vagus nerve influences fHRV's complex properties, the HRV code.

We aimed to identify a comprehensive set of fHRV measures dependent on the intact vagal innervation. We modeled this condition using fetal sheep surgeries, a major homeokinetic disruptor, without or with bilateral cervical vagotomy. We hypothesized that surgical intervention are associated with changes in measures of fHRV and vagotomy alters this association shedding further light onto the contributions of vagal activity fluctuations to the fHRV properties and maintenance of homeokinesis.(Iberall & McCulloch, 1968; Thayer & Sternberg, 2006; Macklem & Seely, 2010) We use this experimental approach to identify a subset of fHRV measures that characterise vagal contributions, the vagus code.(Kwan *et al.*, 2016) We present a methodological tool box to do that borrowing from methodologies used in genomics.

**Methods**

Animal care for the *in vivo* approach followed the guidelines of the Canadian Council on Animal Care and the approval by the University of Montreal Council on Animal Care (protocol #10-Rech-1560).

*Anesthesia and surgical procedure*

We instrumented pregnant time-dated ewes at 126 days of gestation (dGA, ~0.86 gestation) with arterial, venous and amniotic catheters and ECG electrodes as described before.(Burns *et al.*, 2015) For anaesthesia, the ewes were premedicated with acepromazine (Atravet 10 mg/mL) 2 mg intravenously; 30 minutes later, the animals were induced with a combination of diazepam (Diazepam 5 mg/mL) 20 mg, ketamine (Ketalar 100 mg/mL) 4-5 mg/kg and Propofol (Propofol 10 mg/mL) 0.5 to 1 mg/kg. An endotracheal tube was inserted orally and ewes were ventilated in order to maintain a $P_aCO_2$ of 35 to 45 mmHg. Balanced poly-ionic solution was administered at 10 mL/kg for the first hour of general anesthesia and then reduced to 5 mL/kg/h.

Ovine singleton fetuses of mixed breed were surgically instrumented with sterile technique under general anesthesia (both ewe and fetus). In case of twin pregnancy, the larger fetus was chosen based on palpating and estimating the intertemporal diameter. The total duration of the procedure was carried out in about 2 hours. Antibiotics were administered to the mother intravenously (Trimethoprim sulfadoxine 5 mg/kg) as well as to the fetus intravenously and into the amniotic cavity (ampicillin 250 mg). Amniotic fluid lost during surgery was replaced with warm saline. The catheters exteriorized through the maternal flank were secured to the back of the ewe in a plastic pouch. For the duration of the experiment the ewe was returned to the metabolic cage, where she could stand, lie and eat ad libitum while we monitored the non-anesthetized fetus without sedating the mother. During postoperative recovery antibiotic administration was continued for 3 days. Arterial blood was sampled for evaluation of maternal and fetal condition and catheters were flushed with heparinized saline to maintain patency.

During surgery (once the first fetal arterial brachial catheter was in place and before returning the fetus to the uterus) a 3 mL fetal arterial blood sample was taken for blood gas, lactate and base excess (BE) (ABL800Flex, Radiometer) and cytokine measurements. Precordial fetal ECG was placed at first step of fetal instrumentation and recorded for the duration of surgery.

Bilateral cervical vagotomy was performed in selected sheep fetuses as follows. The head was extended and the neck was kept straight and stable while approaching and exposing the vagal nerves. We then performed a bilateral cervical vagotomy while recording fetal ECG. The skin was closed in a continuous pattern. The animals not subjected to vagotomy underwent a sham neck surgery with all steps except vagotomy (sham group).

*Experimental protocol*

Postoperatively, all animals were allowed 3 days to recover before starting the experiments. On these 3 days, at 9.00 am 3 mL arterial plasma sample were taken for blood gases analysis. For the purposes of this study, each experiment was completed at 9.00 am with a 1 h baseline measurement followed by the respective intervention as reported elsewhere. (Burns *et al.*, 2015; Durosier *et al.*, 2015) After the +54 hours (Day 3) sampling, the animals were sacrificed. The 72 h post-surgery, i.e., 1 h baseline, time point is reasonable to consider as a post-surgical recovery since the pharmacological effects of the anaesthetic with respect to the cardiovascular system would have worn off by then. However, there may still be an effect caused by the anaesthesia on the immune system which would be difficult to identify. We have previously shown that this surgical procedure does not result in systemic fetal inflammation.(Burns *et al.*, 2015)

*FHRV analysis*

To derive fHRV, fetal ECG recordings were studied over 5 min periods, starting 30 min prior to vagotomy and 15 min after vagotomy to ensure stable recordings. Thus, fHRV was analyzed at three time points (cf. experimental diagram in Fig. 1): 1) before and 2) after vagotomy during surgery or at the equivalent point in time of surgery in sham animals followed 3) by a baseline measurement 72 h after surgery. The CIMVA (continuous individualized multiorgan variability analysis) software engine was used to develop comprehensive continuous fHRV measures analysis.(Herry *et al.*, 2013) Briefly, the analysis was as follows. First individual heartbeats were identified, using commonly used algorithms a time series of R-peak to R-peak intervals (RRI) was formed. A thorough automated assessment was performed of the quality of the underlying physiological waveform signal and derived physiological events time series, based on the morphology of the ECG/respiratory waveforms, the level of noise/artifacts (including baseline drifts, spikes, and movement artifacts), the proportion of disconnected/saturated periods, and the presence of non-stationarity. Using the cleaned RRI time series, the signal complexity and degree of variability was assessed, through a moving window analysis, whereby a window of fixed duration (5 minutes) is shifted in time across the entire duration of the event time series. A comprehensive set of variability metrics are calculated within each window. The abbreviation list of all fHRV measures we calculated is provided in the Table S1. This analysis is performed iteratively and repetitively, measuring variability over time intervals. Only high-quality (as per the automated quality assessment) 5-minute-windows were retained for the subsequent analysis. For each animal, at each time point and for each fHRV measure, an average of the 30 preceding minutes was calculated.

Thus, the output of the fHRV analysis was a matrix of measures linked to the timing of blood sampling for arterial blood gases and pH, which allowed us to assess the temporal relation of multiple fHRV measures to manipulations of the fetal innate immune system's response to endotoxin and the potential clinical value of linear and nonlinear fHRV measures for monitoring fetal inflammatory response.

*Statistical analysis*

Generalized estimating equations (GEE) modeling was used to assess the effects of treatment (sham, vagotomy) while accounting for repeated measurements on fetal blood gases, lactate and base excess (BE). We used a linear scale response model with time and treatment as predicting factors to assess their interactions using maximum likelihood estimate and Type III analysis with Wald Chi-square statistics. Table S1 presents the complete list of fHRV measures studied using CIMVA approach and their classification into signal-analytical domains and aspect of variability measured (degree or complexity of fHRV). As the CIMVA panel considered 46 HRV measures, each compared twice, a test-wise correction for multiple comparisons was made after performing two-sided Wilcoxon rank sum test at 5% significance level corrected with a false-discovery rate method of Benjamini–Hochberg–Yekutieli. To assess a median difference between the sham and vagotomy groups, we compared the three time points (surgery start, surgery end and 72h recovery) expressed as gain scores, i.e., pairwise difference between the time points. SPSS Version 21 was used for these analyses (IBM SPSS Statistics, IBM Corporation, Armonk, NY). The results are presented as Mean±SD, for P<0.05.

Supplementary data are linked in text and can also be accessed on GitHub (https://github.com/martinfrasch/vagus_HRV_code).

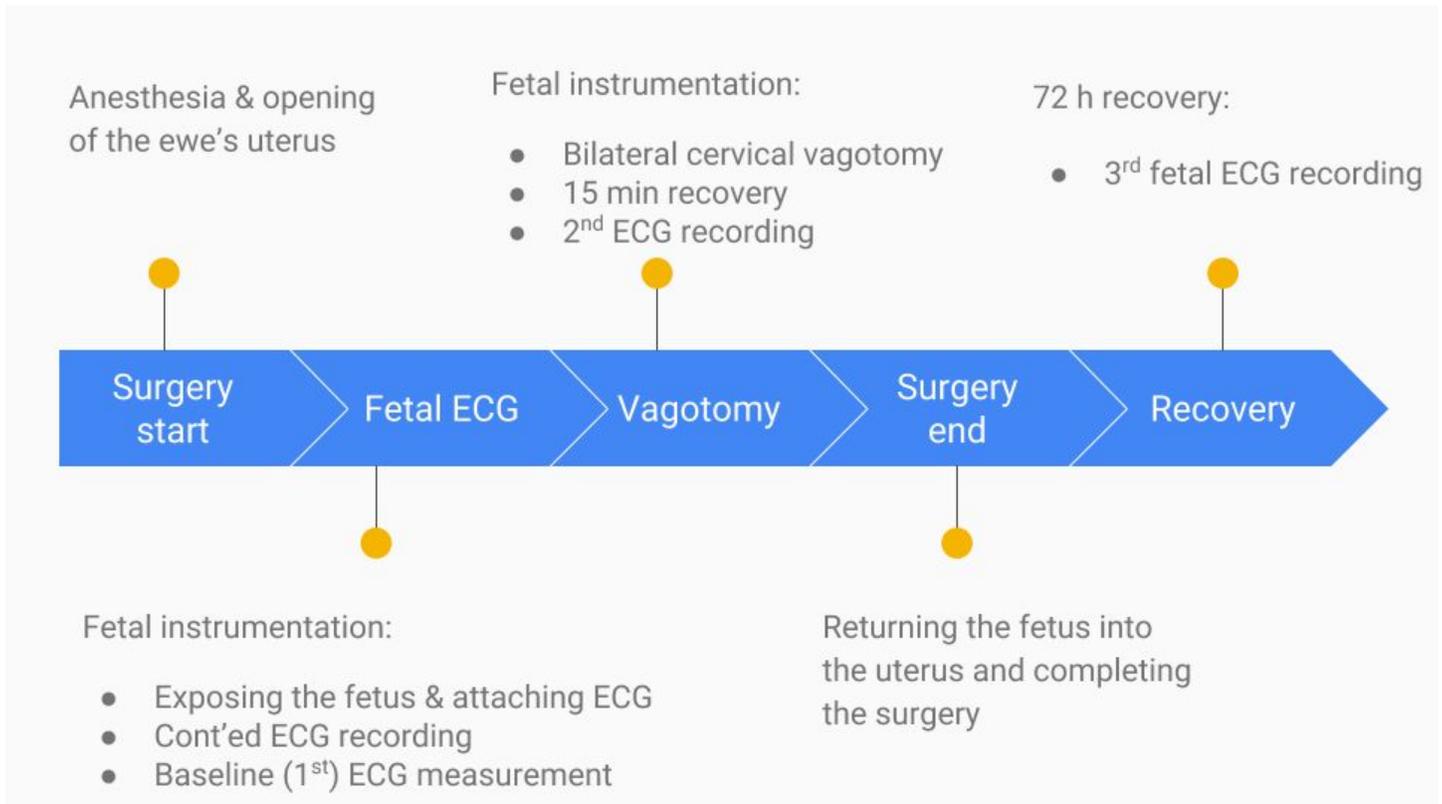

**Figure 1.** Experimental approach. ECG was analysed from 15 min segments taken 15 min prior to and 15 min after vagotomy during the surgical instrumentation as well as 72 hours after surgical recovery.

**Results**

*Fetal arterial blood gas and acid-base status*

We found significant effect of the main term time for all variables (observations prior and after surgery and 72 h recovery, P=0.000, Table 1). We found significant effect of the main term vagotomy for pH, $O_2$Sat, BE (P=0.000, P=0.010, P=0.000, respectively) and no interaction between time and vagotomy (time*vagotomy, P=0.210, P=0.803, P=0.054). There was no significant effect for $pCO_2$, $pO_2$ and lactate for vagotomy (P=0.685, P=0.672, P=0.335); for $pCO_2$ there was a significant interaction (P=0.024). For $pO_2$ and lactate the interaction term was not significant (P=0.558 and P=0.856). In summary, pH, $O_2$Sat and BE showed similar recovery dynamics, but were higher in vagotomy than sham animals. That is, vagotomy group animals were slightly less acidotic than the sham animals.

*Effect of vagotomy on post-operative recovery of fHRV*

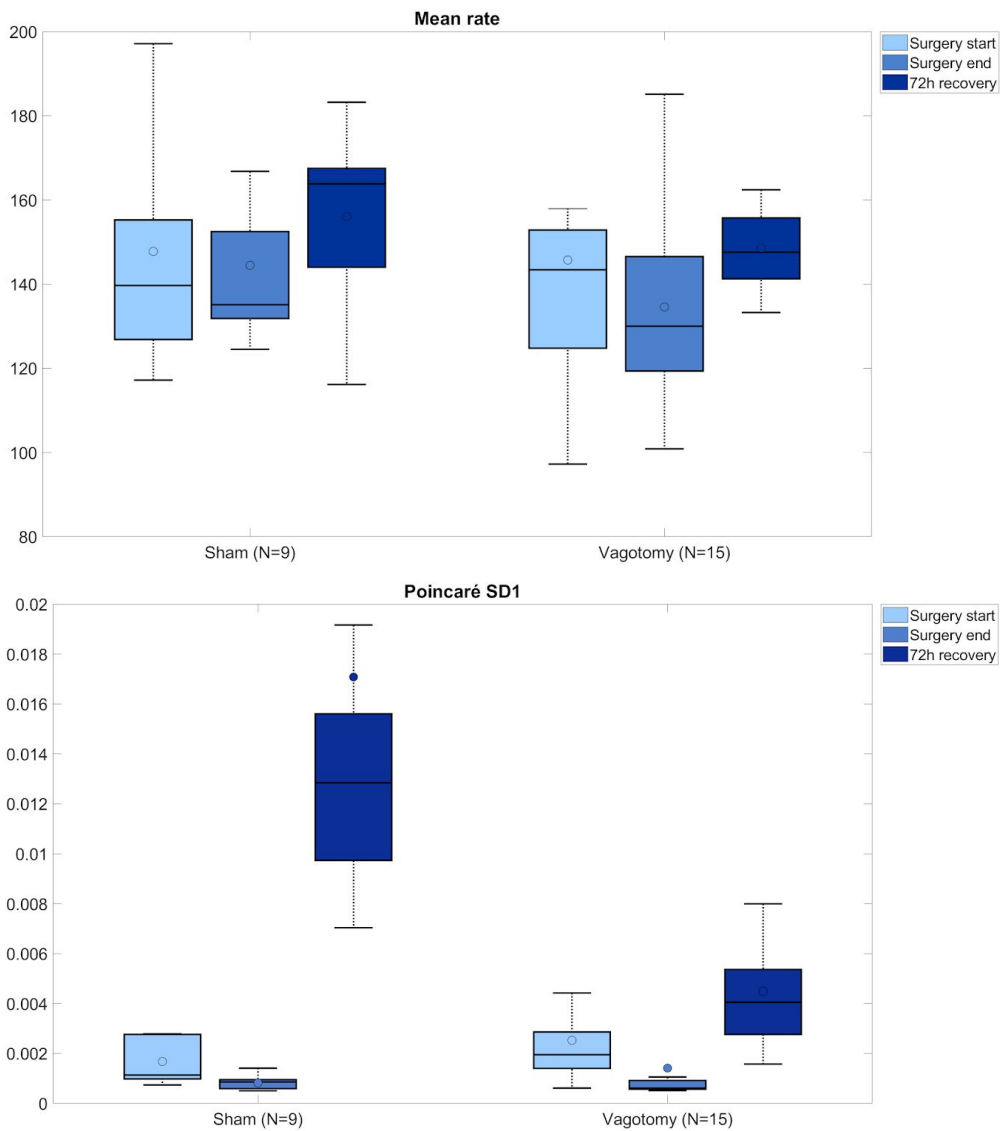

**Figure 2.** Fetal heart rate variability (fHRV) measures during and after surgery without (sham) and with vagotomy. Note no change in FHR in both groups, while a representative HRV measure Poincare SD1 (equivalent to RMSSD) capturing short-term time scale HRV fluctuations shows a recovery at 72h post-surgery in the sham, but not in vagotomy group. The comprehensive review of all changes in fHRV is presented in Fig. 3 (patterns).

FHR remained unchanged within each group averaging 145 ± 25 bpm throughout the entire observation period (Fig. 2).

At the onset of surgery, both groups showed no differences in fHRV properties. At surgery end compared to surgery onset, we observed no change in fHRV measures in either sham or vagotomy animals. In other words, we saw no acute effects of surgery on fHRV regardless the vagotomy.

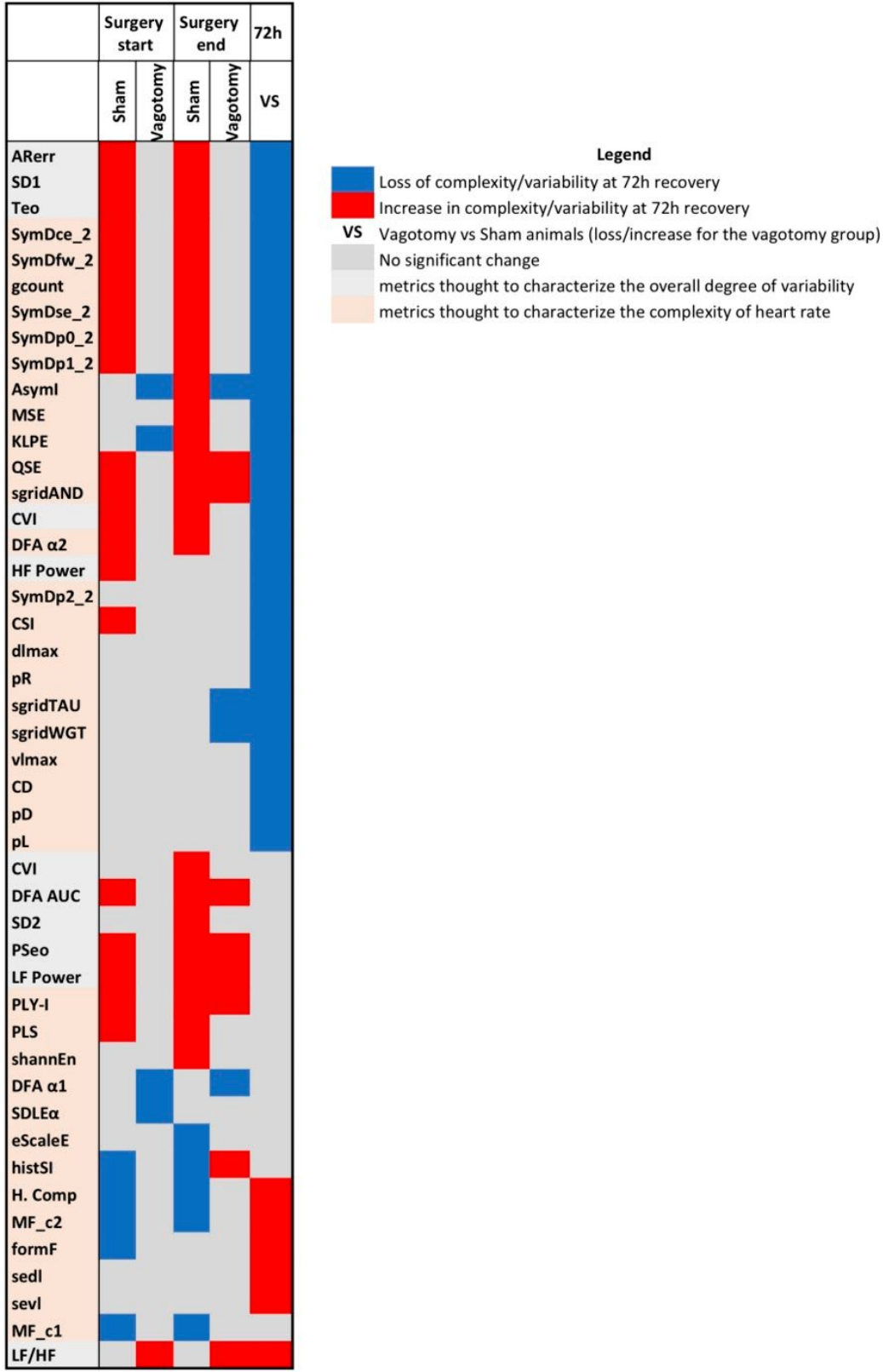

**Figure 3. Comparison of HRV at 72h recovery to surgery start and end.** Loss (blue) or increase (red) in variability or complexity for HRV measures that exhibited significant changes at 72h recovery vs surgery start or surgery end time points. Grey cells represent non-significant changes. See Table S1 for the abbreviations legend of the HRV measures.

To visualize the changes that fHRV experiences with surgical perturbation without or with vagal denervation, as first step, we used a heat map, grouping measures based on whether they increase or decrease in complexity and/or overall degree of variability across time groups and between sham and vagotomy groups.(Sorani *et al.*, 2007; Herry *et al.*, 2016) We considered 46 fHRV measures (Fig. 3).

At 72 h recovery compared to surgery onset, we observed group-specific fHRV changes. For sham animals, 25 out of 46 fHRV measures were changed by surgery with 20 measures showing increased complexity or variability and five measures showing a loss of complexity or variability. For vagotomy animals, we observed a change in 5 fHRV measures of which 1 showed increased complexity or variability and 4 showed a loss of complexity or variability.

When comparing to surgery end, at 72 h recovery we saw a similar pattern with 29 fHRV measures changed in sham animals of which 24 measures showed increased complexity or variability, while 5 showed a loss of complexity or variability. In vagotomy group animals, 12 fHRV measures changed of which 8 measures showed increased complexity or variability, while 4 measures showed a loss of complexity of variability.

When comparing vagotomy versus sham group animals, most fHRV measures differed (33 out of 46) with 6 measures showing increased complexity or variability in vagotomy animals at 72 h recovery compared to sham animals and 27 measures showing loss of complexity or variability.

Next, to identify a core subset of fHRV measures reflecting HRV spontaneous activity, among the tested fHRV measures, we determined how many of the fHRV measures altered at 72 h recovery were specific to surgery with or without vagotomy. Fig. 4 depicts the sets of fHRV measures specific or common to the effect of surgery (taken as 72 h recovery versus surgery start time) in both groups and the effect of vagotomy. To build the Venn diagrams we defined the physiological meaning of recovery in terms of fHRV changes as absence of significant difference at 72 h recovery versus surgery start time (Fig. 4A); we included those measures that rebounded significantly *above* the surgery start values, since variability and complexity increased, commonly associated with healthy states (Fig. 4A). We considered "the lack of recovery" when the opposite condition was fulfilled, *i.e.*, values significantly *below* the surgery start time point (Fig. 4B).

Figure 4A shows 42 and 41 measures that recovered in vagotomy and sham groups, respectively, with 37 measures being in common for both groups. Figure 4B shows two non-overlapping subsets of four measures that did not recover in either group.

While the LF/HF ratio increased, LF power and HF power were not significantly changed and therefore it is not clear whether the ratio indicates a true increase in variability with respect to our assumption and as a result LF/HF ratio was pooled with the 4 other measures that show a loss of complexity or variability.

Vagotomy limits the fHRV recovery and identifies a subset of four fHRV measures that are affected by vagotomy: AsymI, DFA $α_1$, KLPE and SDLE α.

Interestingly, the 20 fHRV measures that bounced back above values at the surgery start time point did not return to pre-surgery values during the subsequent 49 hours of control conditions (previously reported (Durosier *et al.*, 2015)) possibly indicating that the impact of surgery on HRV is more long-lasting than previously thought.

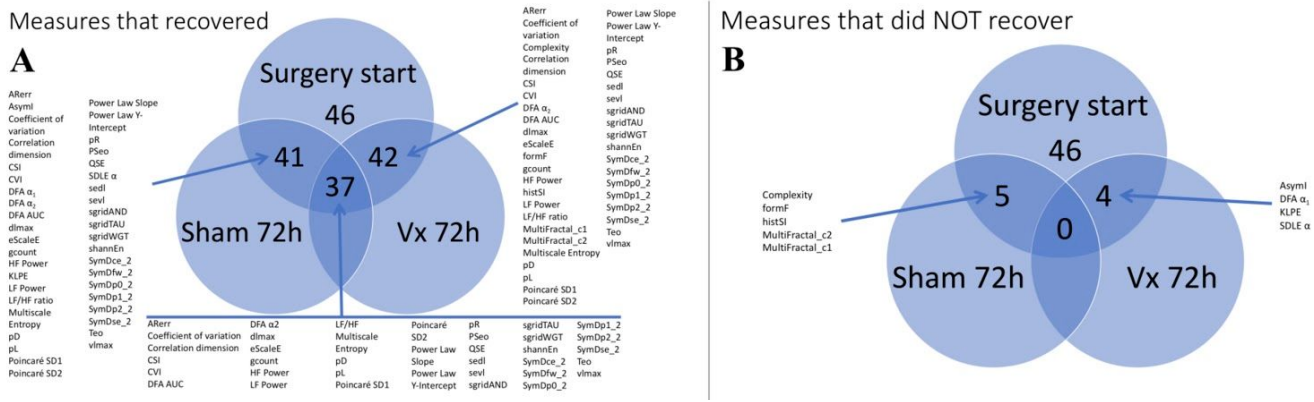

**Figure 4.** Venn diagram representation of changes to fHRV code (represented by 46 fHRV measures) under recovery from surgery and the influence of vagotomy. See Table S1 for the abbreviations legend of the HRV measures.

**Discussion**

Little is known about the effects of fetal surgery on fHRV as a marker of fetal health and recovery from a major insult. We identified subsets of fHRV measures that can now be validated as fetal postsurgical recovery markers. The relative difference between 72h and start or end of surgery is overall smaller in vagotomy animals than in sham animals. Combined with findings from Figure 4, we conclude that vagal denervation results in slower recovery from surgery. Our findings suggest that drugs influencing fetal vagal cholinergic signaling delay the HRV recovery and identify a set of four fHRV measures that reflect that effect: AsymI, DFA $\alpha_1$, KLPE and SDLE α.

While our study does not permit inference onto the categories of sympathetic or cardiovascular HRV contributions, we identified subsets of fHRV measures depending on intact vagus signaling. In following we attempt two secondary analyses on existing fHRV data sets from related experiments (same gestational age): a) contribution of the vagus nerve signaling to fHRV inflammatory signature and b) the comparison to the recently identified intrinsic HRV (iHRV) signature.

*a) comparison to systemic and organ-specific fHRV inflammatory signatures*

Via the vagus nerve, brain senses systemic and organ-specific levels of inflammatory cytokines.(Garzoni *et al.*, 2013; Zanos *et al.*, 2018) We showed that this process is reflected in specific subsets of HRV measures. In previous studies conducted in this cohort's animals with intact vagus nerves (Durosier *et al.*, 2015; Herry *et al.*, 2016; Liu *et al.*, 2016), we identified a systemic LPS exposure response-specific fHRV signature as well gut- and brain-specific HRV signatures (Table S2). Here, we compared previously identified fHRV measures to those shown in this work to identify whether they change uniquely due to vagotomy or persist despite the vagotomy. We hypothesized that there would be minimal overlap between the vagotomy group in the present study and the inflammatory signatures, because the latter represent a signaling process dependent on an intact vagus nerve.(Garzoni *et al.*, 2013)

To test this hypothesis, we examined the fHRV signature of vagotomy animals at 72 hours of recovery to avoid contamination with acute post-surgical recovery effects. We found that the multiscale time irreversibility asymmetry index (AsymI) of fHRV, present in the systemic inflammatory index, did not recover in vagotomy group underscoring the connection of AsymI to vagus nerve signaling via the cholinergic anti-inflammatory pathway (cf. Fig. 4B).(Garzoni *et al.*, 2013; Durosier *et al.*, 2015) DFA $\alpha_1$ did not recover in vagotomy animals as a measure shared with the organ-specific inflammatory index (for terminal ileum's M2 (CD206+) macrophages and Iba1-positive brain microglia - cf. Table S2).

For brain-specific inflammation signature, we found no further fHRV measures that reflected vagus nerve activity. Three out of five fHRV measures that did not recover in sham animals were also found in the brain inflammation signature: form factor (formF), multifractal spectrum cumulant of the first order (MultiFractal_c1) and similarity index of distributions (histSI). This supports our above proposed notion of HRV spontaneous activity (Fig. 4B). The present findings suggest that is represented by formF, MultiFractal_c1 and histSI measures. Such activity can be perturbed either by brain inflammation or by the homeokinetic adaptations taking place after surgery.

*b) Comparison to iHRV signature*

In a recent study of fetal sheep heart *ex vivo*, *i.e.*, devoid of any innervation, we identified a signature of iHRV activity in response to *in vivo* hypoxia.(Frasch *et al.*, 2018*b*) The iHRV signature is comprised of two recurrence quantification metrics (dlmax, pL), a scale-dependent Lyapunov exponent metric (SDLEα) and sgridAND, a grid transformation metric. Here we compared these iHRV measures to the fHRV measures reflecting lack of vagal innervation (Fig. 4B). We hypothesized that there would be an overlap in the fHRV measures in both groups, since they both represent iHRV, although other neural and non-neural extracardiac influences we could not control for in the present study likely played a role in the *in vivo* recorded fHRV with vagal denervation. Consequently, the fHRV measures that do not overlap may represent sympathetic or other, non-neural, influences on fHRV in the vagotomy group.

Figure 4A identifies a subset of fHRV measures recovering regardless of vagotomy that also belong to the above iHRV signature: dlmax, pL, sgridAND. This finding strengthens the previously reported iHRV signature. For SDLE α which did not recover in vagotomized animals, we propose that *in vivo* formation of hypoxia memory in iHRV requires vagus-nerve-mediated imprinting on the iHRV signature. The behavior of these four fHRV measures should be studied longitudinally *in vivo* and *ex vivo* in future physiological studies.

*General discussion*

We propose to conceptualize the ensemble behavior of the multitude of the fHRV measures over time in terms of spontaneous and evoked activity reflected in HRV code, a time series of multi-dimensional dynamic patterns. We report here that the HRV code of the spontaneous activity overlaps considerably with the brain (but not systemic or gut's) inflammation signature (cf. Fig. 4B). It may also be influenced by the regulatory activities of the cardiovascular and autonomic nervous systems and, more generally, suprabulbar brain's efferent signaling via these systems. HRV's response to a general insult of surgery can explain the overlap with the features of the brain inflammation signature of the HRV code.

The complexity of HRV pattern is highlighted by the finding that the five fHRV measures that did not recover in sham animals are distinct from those in the vagotomized animals (Fig. 4B). This supports the concept that we can identify specific HRV signatures of a general perturbation in the spontaneous activity as well as the HRV signature reflecting uniquely the withdrawal of vagal innervation.

We identified a set of fHRV measures characterizing recovery from a surgery, a major injurious stimulus of general nature perturbing homeokinesis. The removal of vagus innervation altered the chronic recovery of the fHRV code represented by these measures. In other words, vagotomy perturbs such spontaneous activity further reflecting underlying HRV fluctuations which represent a proxy to the vagus code embedded in fHRV.(Kwan *et al.*, 2016)

In sham animals, most fHRV measures recovered with complexity and variability increasing at 72h when compared to pre or post-surgery time points. However, 30-35% of fHRV measures showed a loss of complexity. If we are to account for collinearity in the measures, this subset of "dissenting" measures constituted <10% (five measures in total). These "dissenting" fHRV measures persisted for one week after surgery and might serve as "memory" biomarkers of fetal surgery, although this would need to be tested further during longer observation periods. Another possible explanation might be the simplified interpretation of increase/loss of complexity/variability. For instance, the lower values of Hjorth's Complexity and Form Factor (formF) generally indicate smoother (less complex) signals, but paradoxically can also relate to signals with a higher fractal dimension, suggesting that the they characterize a different type of complexity compared to other metrics. Similarly, multifractal spectrum cumulants of the first and second order (Multifractal_c1, Multifractal_c2) are related to the maximum and width of the multifractal spectrum, respectively, and therefore characterize local regularity in the signal rather than complexity per se. The embedded Scaling exponent (eScaleE) exhibited a very tight range in our data, so differentiating between groups with our relatively small sample size was difficult. Lastly, the similarity index (histSI) is a similarity measure between consecutive time segments and smaller values generally indicate segments are more dissimilar, although this does not always translate to higher intrinsic variability as that measure is affected by low frequency components.(Huang *et al.*, 2008)

Vagotomized fetuses showed an overall pattern of chronic loss of complexity and overall degree of variability. We observed eight dissenting metrics that were significantly higher in vagotomy animals at 72h compared with sham animals, most of them consistent with the previously discussed dissenting metrics with the exception of sedl and sevl (Shannon entropy of diagonal and vertical lines, respectively, from the recurrence plots) and LF/HF ratio. The increase in the Shannon entropy of diagonal and vertical lines could be due to noise effects, while the LF/HF ratio increase is most likely due to a drop in HF power.

For vagotomy animals, we observed some recovery of complexity or variability, but it was reflected by a different set of fHRV measures than for the sham animals. Comparing both groups at 72 hours, our data shows that vagotomy hindered the recovery of complexity and variability after surgery. That is in line with the notion that vagus nerve is homeokinetic in nature, hence vagus' withdrawal would be expected to interfere with the surgical recovery.(Thayer & Sternberg, 2006; Frasch *et al.*, 2009) Here we identified a set of fHRV measures that reflect this effect, most of which are not yet conventionally used as biomarkers of vagal modulation of HRV.

When comparing with previous work on systemic and organ-specific fHRV inflammatory signatures, our findings indicate that AsymI might be uniquely tied to the vagus nerve signaling via the cholinergic anti-inflammatory pathway. This result is supported by the studies in septic neonates where AsymI is part of the HRV-derived inflammatory index to predict early onset of neonatal sepsis.(Lake *et al.*, 2014) To our knowledge, the present study is a first mechanistic demonstration of the connection between AsymI and vagus nerve signaling linking it to the cholinergic anti-inflammatory pathway.(Andersson & Tracey, 2012) Furthermore, both AsymI and DFA $\alpha_1$ HRV measures appear to be components of the vagus code dynamics in response to inflammation and can serve as biomarkers of inflammation and organ-specific vagus code signatures. Conversely, we observed the presence of HRV measures in the organ-specific inflammation signatures that do not map uniquely onto the contribution of the vagus nerve activity. Such behavior is to be expected and yields the more general HRV code, as the sensing and control of inflammation represents a complex homeokinetic process with nonlinear contributions from multiple physiological systems. We suggest that HRV code in general and vagus code in particular ultimately represent an emergent property reflecting the underlying communication via the vagus nerve and other complex oscillatory influences. As reviewed elsewhere (Kwan *et al.*, 2016) and demonstrated in several recent studies (Steinberg *et al.*, 2016; Datta-Chaudhuri *et al.*, 2018), the the organ specific vagus code can be captured directly via vagus nerve electroneurogram (VENG). The precise delineation of the relationships between VENG properties and HRV code properties is subject of future studies.

*In utero* hypoxia likely reduces vagal tone in favor of sympathetic activity.(Morrison, 2008) If this imbalance is imprinted with a "1:1" transfer function onto iHRV, we would expect a similar pattern with increased percentage of Laminarity (pL) and maximum diagonal line (dlmax) in the Recurrence plots in *ex vivo* hypoxic hearts, but the opposite was found consistently. This suggests that the drop in pL and dlmax may reflect "true" iHRV dynamics or the *ex vivo* memory of hypoxia is not retained in a linear fashion.

The present study has several limitations. First, the time points for assessment of the acute fHRV changes during surgery as a function of vagotomy were at 30 min relatively close to each other. However, we deliberately targeted stable pre- and post-vagotomy segments of the surgery which both shared the same anesthetic regime and surgical procedures, including a sham vagotomy in the sham experimental group. Second, we conducted a number of comparisons for fHRV measures in this study and adjusted for the multiple comparisons within each fHRV measure for each time point. As recommended by Rothman(Rothman, 1990), we did not adjust for the overall number of comparisons across all possible tests.

Second, we assumed that our set of HRV measures represented an adequate characterization of the variability of the complex cardiovascular system and attempted to reveal specific HRV signatures triggered by physiological perturbations, inflammation and iHRV/hypoxia. However, it is possible that these HRV signatures we identified underestimate the true dimensionality of the process, even if they correlate or match well with some of its dynamic features. Future studies should attempt to approach this problem directly, rather than inversely, *e.g.*, by recording fetal VENG and sympathetic nerve activity and studying its properties in conjunction with changes in HRV code. We are preparing reports on the respective methodological approach and first physiological findings.(Frank *et al.*, 2018; Frasch *et al.*, 2018*a*)

We consider it beyond the scope of the present study to elucidate the reasons for specific changes in each fHRV measure and leave it to future work. We hope that our current work raises awareness of paradoxical responses in fHRV measures. This should be considered during interpretation of the effects of reduced vagal modulation in other experiments. Our present work highlights the difficulty in summarizing fHRV changes as being more/less complex or suppressed/increased variability.(Frasch, 2018) The various HRV measures characterize different, sometimes conflicting, aspects of the signal and the dynamics of the underlying system. It is for this reason that we introduced the terms "HRV code" and "vagus code". The search for physiological meaning is ongoing. This highlights the continued need for collaborative transdisciplinary efforts to refine further the multi-dimensional / multi - domain HRV concept toward the formulation of a robust mathematically-founded biological HRV code. We hope the present study contributes to this development.

**Acknowledgements**
Supported by CIHR, FRQS, Molly Towell Perinatal Research Foundation.

**Conflict of interest statement**

AJE Seely founded Therapeutic Monitoring Systems in order to commercialize patented Continuous Individualized Multiorgan Variability Analysis (CIMVA) technology, with the objective of delivering variability-directed clinical decision support to improve quality and efficiency of care. CL Herry is a patent holder related to waveform quality assessment necessary for variability analysis. MG Frasch holds a PCT on fetal ECG monitoring. All other authors declare that they have no conflict of interest.

Table 1. Impact of vagotomy on fetal arterial blood gas and acid-base status during surgery start, surgery end and after 72 hours of recovery.

| | Condition | pH | | | $paCO_2$ | | | $paO_2$ | | | $O_2Sat$ | | | Lactate | | | Base Excess | | |
|---|---|---|---|---|---|---|---|---|---|---|---|---|---|---|---|---|---|---|---|
| **Sham** | Surgery start | 7.25 | ± | 0.05 | 58.6 | ± | 10.1 | 29.1 | ± | 7.5 | 70 | ± | 14.1 | 2.3 | ± | 0.8 | -1.6 | ± | 3 |
| | Surgery end | 7.26 | ± | 0.04 | 57.9 | ± | 8.1 | 26.3 | ± | 5.9 | 65.2 | ± | 10.8 | 2.3 | ± | 0.6 | -1.1 | ± | 3.3 |
| | 72h recovery | 7.35 | ± | 0.03 | 51.4 | ± | 5 | 21.6 | ± | 13.2 | 48.8 | ± | 15.5 | 1.9 | ± | 1.1 | 2.4 | ± | 2.4 |
| **Vagotomy** | Surgery start | 7.29 | ± | 0.04 | 55.5 | ± | 8 | 31.2 | ± | 6.6 | 77 | ± | 7.7 | 2.5 | ± | 0.6 | -0.3 | ± | 2.8 |
| | Surgery end | 7.28 | ± | 0.04 | 60 | ± | 8.3 | 26.3 | ± | 3.5 | 67.5 | ± | 8.5 | 2.5 | ± | 0.7 | 1 | ± | 2.3 |
| | 72h recovery | 7.37 | ± | 0.03 | 50.6 | ± | 4.3 | 19.6 | ± | 3.8 | 52.3 | ± | 13.8 | 2 | ± | 0.8 | 3.4 | ± | 2.1 |